\documentclass[12pt]{spieman}  
\usepackage{amsmath,amsfonts,amssymb}
\usepackage{graphicx}
\usepackage{setspace}
\usepackage{tocloft}
\usepackage{subcaption}
\usepackage[left]{lineno} 
\title{Proton radiation hardness of X-ray SOI pixel sensors with pinned depleted diode structure}

\author[a]{Mitsuki~Hayashida}
\author[a,*]{Kouichi~Hagino}
\author[a]{Takayoshi~Kohmura}
\author[a]{Masatoshi~Kitajima}
\author[a]{Keigo~Yarita}
\author[a]{Kenji~Oono}
\author[a]{Kousuke~Negishi}

\author[b]{Takeshi~G.~Tsuru}
\author[c]{Takaaki~Tanaka}
\author[b]{Hiroyuki~Uchida}
\author[b]{Kazuho~Kayama}
\author[b]{Ryota~Kodama}

\author[d]{Koji~Mori}
\author[d]{Ayaki~Takeda}
\author[d]{Yusuke~Nishioka}
\author[d]{Takahiro~Hida}
\author[d]{Masataka~Yukumoto}

\author[e]{Yasuo~Arai}

\author[f]{Ikuo~Kurachi}

\author[g]{Hisashi~Kitamura}

\author[h]{Shoji~Kawahito}
\author[h]{Keita~Yasutomi}

\affil[a]{Department of Physics, School of Science and Technology, Tokyo University of Science, 2641 Yamazaki, Noda, Chiba 278-8510, Japan}
\affil[b]{Department of Physics, Faculty of Science, Kyoto University, Kitashirakawa-Oiwakecho, Sakyo-ku, Kyoto 606-8502, Japan}
\affil[c]{Department of Physics, Konan University, 8-9-1 Okamoto, Higashinada, Kobe, Hyogo 658-8501, Japan}
\affil[d]{Department of Applied Physics, Faculty of Engineering, University of Miyazaki, 1-1 Gakuen-Kibanodai-Nishi, Miyazaki, Miyazaki 889-2192, Japan}
\affil[e]{Institute of Particle and Nuclear Studies, High Energy Accelerator Research Organization (KEK), 1-1 Oho, Tsukuba, Ibaraki 305-0801, Japan}
\affil[f]{D\&S Inc., 774-3-213 Higashiasakawa, Hachioji, Tokyo 193-0834, Japan}
\affil[g]{National Institute of Radiological Sciences, National Institutes for Quantum and Radiological Science and Technology, 4-9-1 Anagawa, Inage-ku, Chiba-City, Chiba, 263-8555, Japan}
\affil[h]{Research Institute  of Electronics, Shizuoka University, Johoku 3-5-1, Naka-ku, Hamamatsu 432-8011, Japan}

\cftpagenumbersoff{figure}
\cftpagenumbersoff{table} 
\begin{document} 
\maketitle


\begin{abstract}X-ray silicon-on-insulator (SOI) pixel sensors, ``XRPIX," are being developed for the next-generation X-ray astronomical satellite, ``FORCE".
The XRPIX are fabricated with the SOI technology, {which makes it possible} to integrate a high-resistivity Si sensor and a low-resistivity Si CMOS circuit. The CMOS circuit in each pixel is equipped with a trigger function, {allowing us to read out outputs only from the pixels with X-ray signals at the timing of X-ray detection.}
This function {thus} realizes high throughput and high time resolution, which enables {to employ anti-coincidence technique for background rejection.}
A new series of XRPIX named XRPIX6E developed with a pinned depleted diode (PDD) structure improves spectral performance by suppressing the interference between the sensor and circuit layers. 
When semiconductor X-ray sensors are used in space, their spectral performance is generally degraded owing to the radiation damage caused by high-energy protons. 
Therefore, before using an XRPIX in space, {it is necessary to evaluate} the extent of degradation of its spectral performance by radiation damage.
Thus, we performed a proton irradiation experiment for XRPIX6E for the first time at HIMAC in the National Institute of Radiological Sciences.
We irradiated {XRPIX6E} with high-energy protons with {a total dose of up to} 40~krad, equivalent to 400 years of irradiation in orbit.
{The 40-krad irradiation degraded the energy resolution of XRPIX6E by $25 \pm 3\%$, yielding an energy resolution of $260.1\pm5.6$~eV at the full width half maximum 
for $5.9~{\rm keV}$ X-rays.}
However, {the value satisfies the requirement for FORCE, 300 eV at 6 keV, even after the irradiation.} 
It was also found that the PDD XRPIX has enhanced radiation hardness compared to previous XRPIX devices.
In addition, we investigated the degradation of the energy resolution; it was shown that the degradation would be due to increasing energy-independent components, e.g., readout noise.  
\end{abstract}


{\noindent \footnotesize\textbf{*}Kouichi Hagino,  \linkable{hagino@rs.tus.ac.jp} }

\begin{spacing}{1} 

\section{Introduction}
\label{sec:intro}  

We have been developing monolithic active pixel sensors, XRPIXs, based on silicon-on-insulator (SOI) CMOS technology as soft X-ray sensors for the FORCE satellite~\cite{Mori16,Tsuru18}. The XRPIX is monolithic, using a bonded wafer of a high-resistivity depleted Si layer for X-ray detection, a SiO$_2$ insulator layer (buried oxide (BOX)), and a low-resistivity Si layer for CMOS circuits. 
The depletion layer can be as thick as {$200$ to $500~{\rm \mu m}$}.
The CMOS circuit in each pixel is equipped with a trigger function, {which allows us to read out outputs only from pixels with X-ray signals at the timing of X-ray detection.}
It realizes high throughput and high time resolution, enabling background rejection with the anti-coincidence technique. 

\begin{figure}[tb]  
 \begin{tabular}{cc}
\begin{minipage}{0.48\hsize}
       \begin{center}
               \includegraphics[width=8cm]{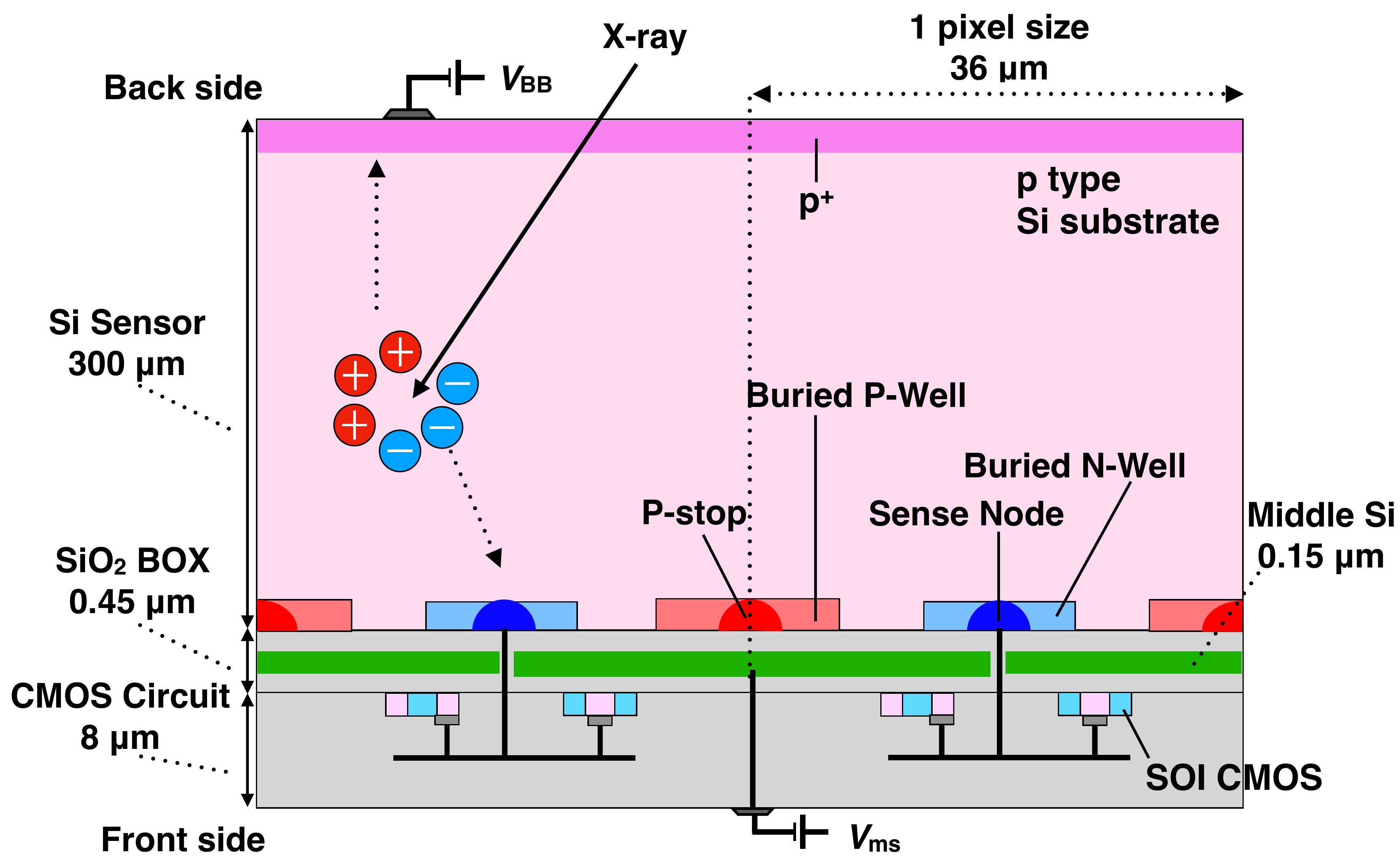} 
        \caption{Cross-sectional view of DSOI XRPIX}
        \label{xr6c}
 \end{center}
\end{minipage}

\hspace{2mm}

\begin{minipage}{0.48\hsize}
       \begin{center}
        \includegraphics[width=8cm]{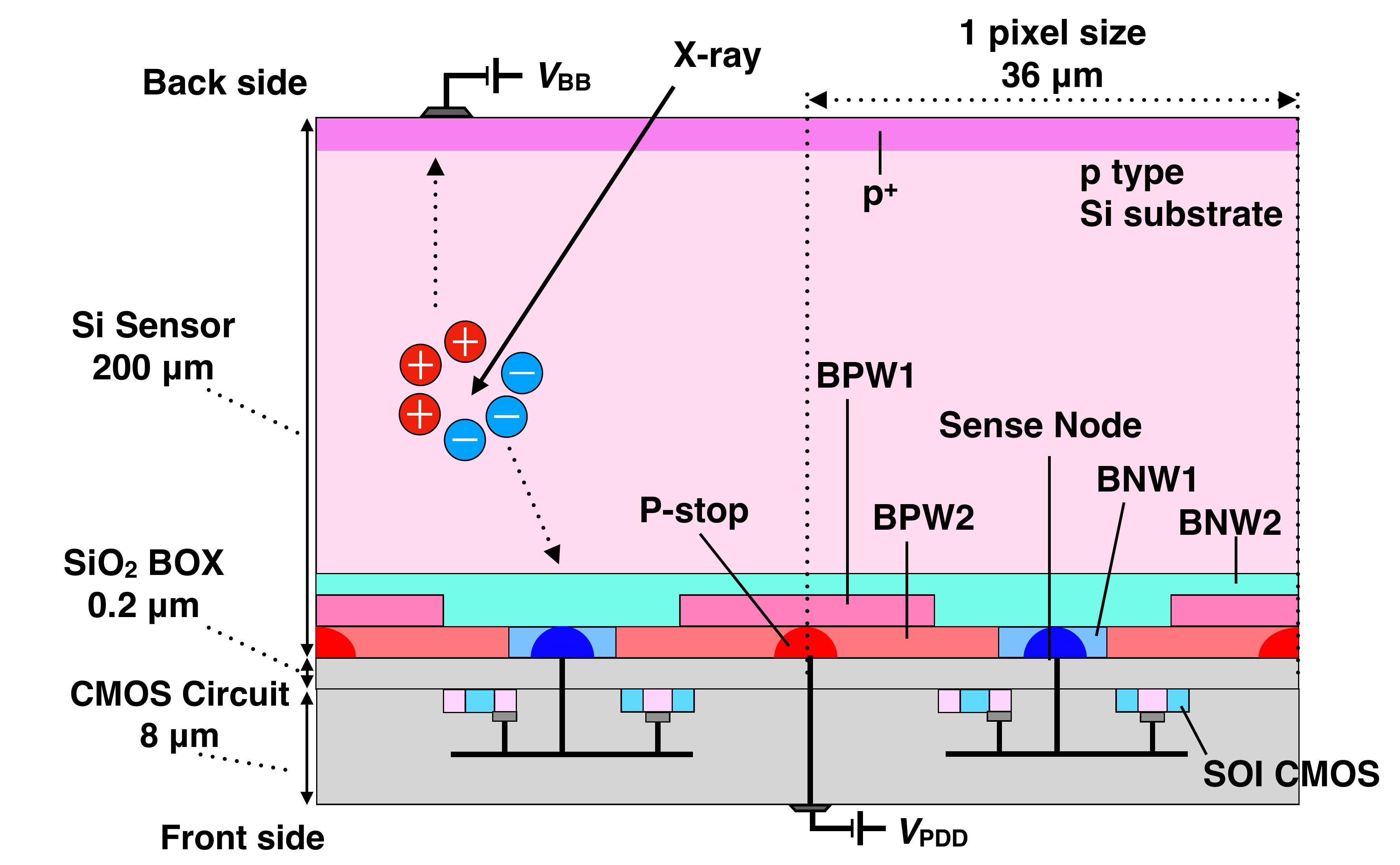} 
        \caption{Cross-sectional view of PDD XRPIX}
        \label{xr6e}
 \end{center}
\end{minipage}
\end{tabular}
 \end{figure}
 
{Improvement of the energy resolution of XRPIX has been one of our major development items.} 
{Testing various XRPIX sensors, we found that the energy resolution is significantly degraded due to the electrical interference between the sensor layer and the trigger circuit.}
{To suppress the interference,} we developed an XRPIX with a double SOI (DSOI) structure {shown in Fig.~\ref{xr6c}~\cite{Takeda20}.}
{In the DSOI XRPIX,} a {low-resistivity} Si layer is inserted in the BOX layer, {to which a negative voltage is applied so that the layer can act as an electrostatic shield.}
However, the XRPIX with the DSOI structure has problems such as charge loss {near the Si--SiO$_2$ interface and large dark current from the interface}~\cite{Hagino19}.
{We then developed an XRPIX with a pinned depleted diode (PDD) structure in order to solve the problems about the interference and the Si--SiO$_2$ interface at the same time}~\cite{Kamehama18,Harada19}.
Fig.~\ref{xr6e} shows the cross-sectional view of the PDD XRPIX.
A sufficiently high doped buried p-well (BPW) is formed at the interface between the sensor and oxide layers and acts as an electrostatic shield.
{The BPW also prevent the dark current and the charge loss at the interface by creating a potential barrier near the interface.}
Consequently, the spectral performance of the PDD XRPIX is improved compared to those of the DSOI XRPIX and previous devices~\cite{Kamehama17,Harada19}.

{The radiation hardness is also an important development item of the XRPIX.}
In general, semiconductor devices, such as X-ray sensors, suffer from radiation effects.
There are three types of radiation effects: displacement damage dose (DDD), total ionizing dose (TID), and single event.
Among them, DDD and TID effects are known to degrade the spectral performance of X-ray sensors~\cite{Mori19,Yarita19,Hagino21,Mori13,Hayashi13}. 
For example, in X-ray charge-coupled devices (CCDs), the lattice defects caused by the DDD effect result in the loss of some of the charges {during transfer}, which consequently degrades the energy resolution. 
In fact, {in-orbit proton bombardment gradually deteriorated the energy resolution of X-ray CCDs onboard the Suzaku satellite}~\cite{Uchiyama13}.
Therefore, {before using X-ray sensors in space,} it is important to evaluate the degradation of the performance, such as dark current, readout noise, and energy resolution.

SOI pixel sensors are known to be sensitive to the TID effect~\cite{Hara19,Yarita19}, which is caused by the oxide-trapped charges and interface traps generated by radiation.
These accumulated charges and traps change the operating characteristics of the CMOS circuits and increase the dark current from the Si--SiO$_2 $ interface.
However, unlike previous XRPIXs, the radiation hardness of the PDD XRPIX {had never been evaluated}.
In addition, the PDD structure is expected to suppress the dark current due to the interface states near the Si--SiO$_2$ interface and compensate the positive potential of oxide-trapped charges.
{Therefore}, to verify the potential of using the PDD XRPIX in the radiation environment in space, we conducted a proton irradiation experiment for it for the first time~\cite{Hayashida2020}.
In this paper, we describe this experiment and report its results.

\section{Proton Irradiation Experiment}

\subsection{Radiation Hardness Required for FORCE}
\label{subsec:RadHardReq}  

The orbit of FORCE has an altitude of $\sim 550~{\rm km}$ and an inclination angle of $\sim 30^\circ$~\cite{Mori16}. 
In this orbit, the sensors suffer radiation damage primarily by the geomagnetically trapped cosmic-ray protons in the South Atlantic Anomaly~\cite{Dachev06,Pierrard14}.
The typical dose rate by the trapped protons on an XRPIX is {approximately $0.1~{\rm krad / year}$~\cite{Yarita19}}.
Because X-ray astronomical satellites typically operate for a few to {$10~ {\rm years}$} in orbit, the total dose {during the mission lifetime} is less than a few kilorads.

One of the required {performances} of an XRPIX for FORCE is an energy resolution of $< 300~{\rm eV}$ at $6~{\rm keV}$ at the full width half maximum (FWHM)~\cite{Tsuru18}.
Therefore, it is necessary to confirm whether an XRPIX will meet the required performance {after irradiation of several-kilorad protons in an on-ground experiment} before launch.

\subsection{Experimental Setup}

\begin{figure}[tb]  
 \begin{tabular}{cc}
\begin{minipage}[t]{0.49\hsize}
       \begin{center}
        \includegraphics[width=\hsize]{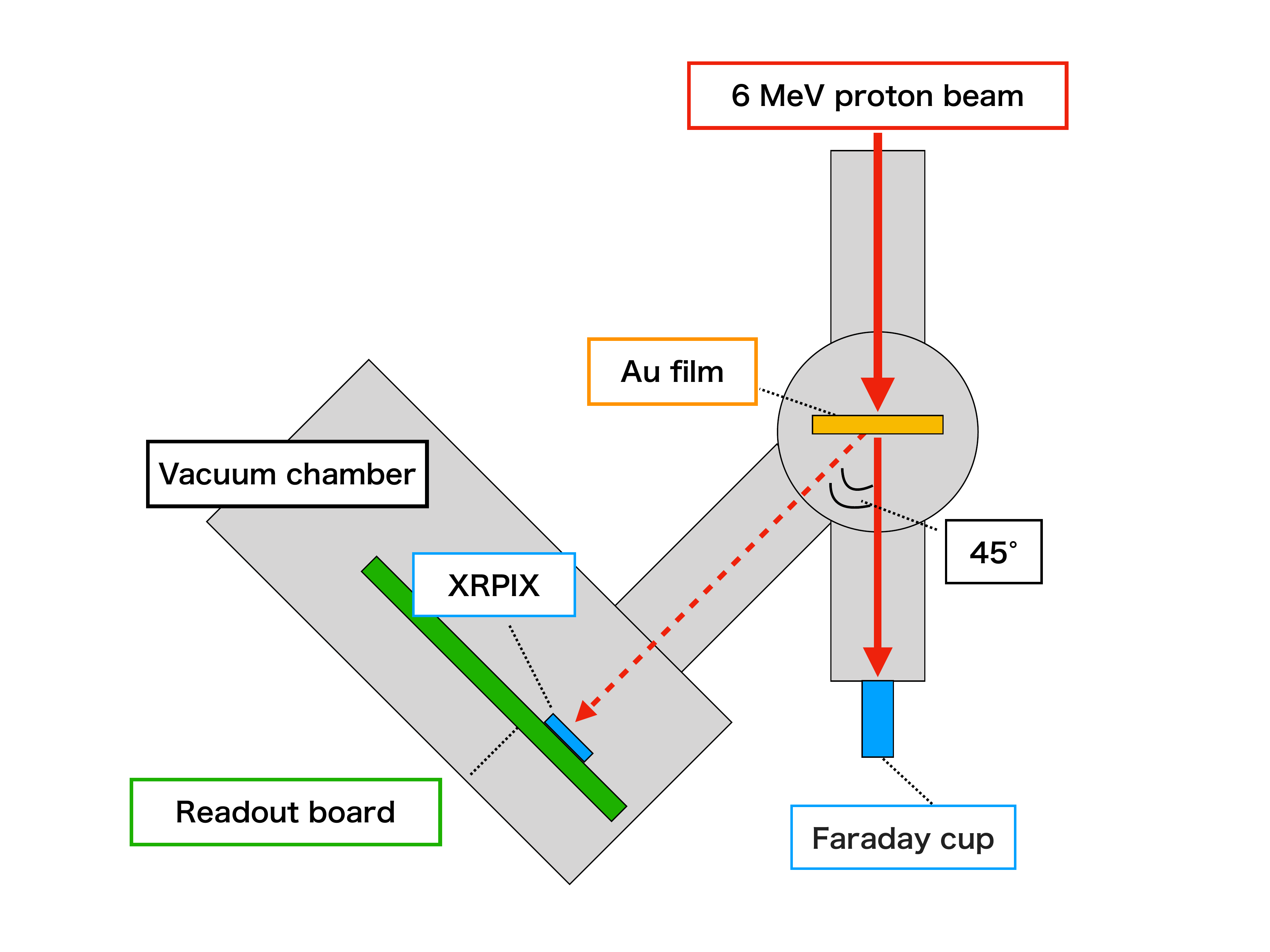}
        \caption[Schematics of top view of 6-MeV proton irradiation experiments]{{Schematics of top view of 6-MeV proton irradiation experiments. In this setup, vacuum chamber is directly connected to beamline.}}
        \label{p_expe6}
 \end{center}
\end{minipage}

\hspace{2mm}

\begin{minipage}[t]{0.49\hsize}
       \begin{center}
        \includegraphics[width=\hsize]{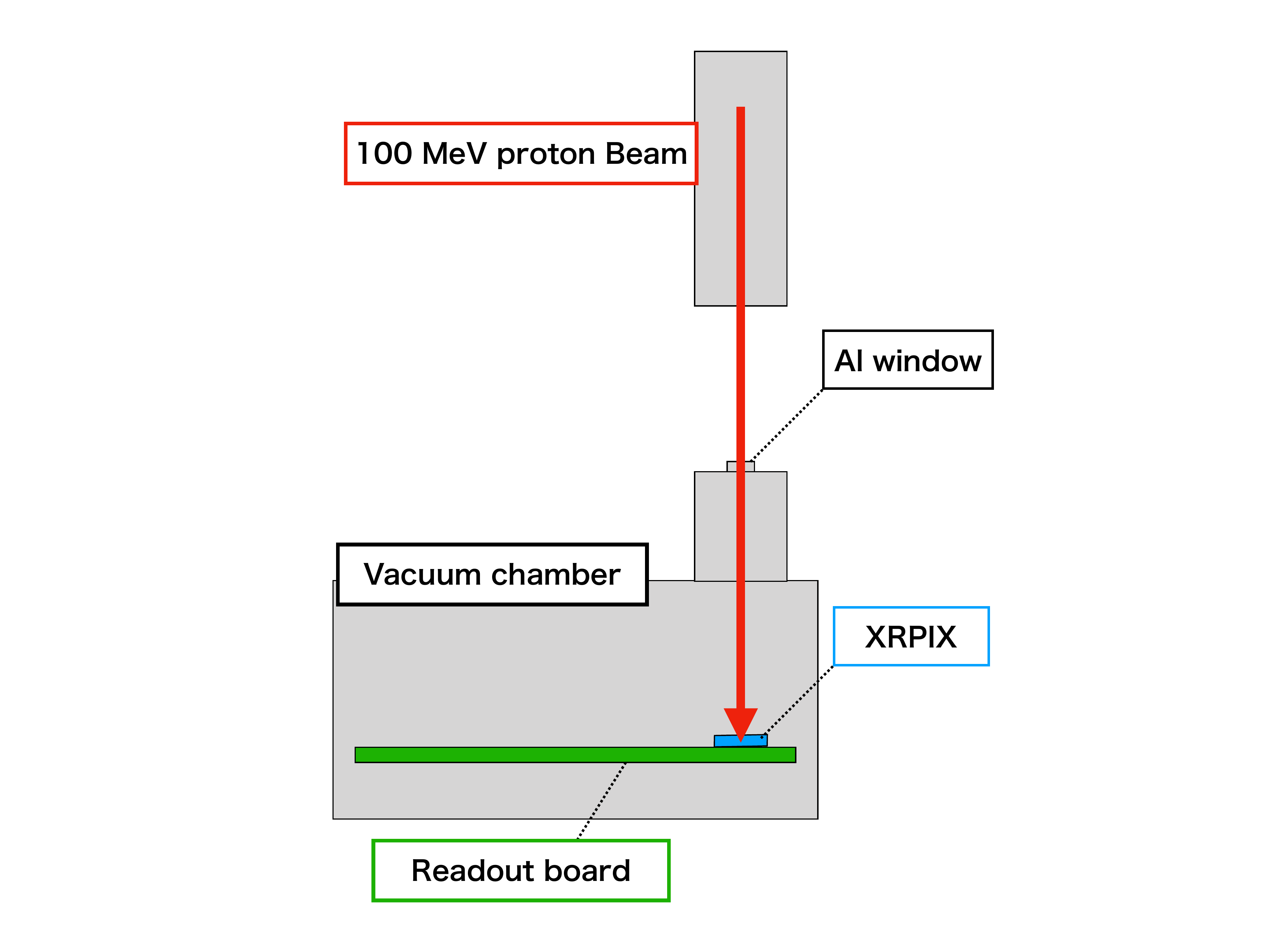}
        \caption[Schematics of top view of 100-MeV proton irradiation experiments]{{Schematics of top view of 100-MeV proton irradiation experiments. In this setup, vacuum chamber is separated from beamline by approximately 1~m, and proton beam is irradiated through Al window.}}
        \label{p_expe100}
 \end{center}
\end{minipage}
\end{tabular}
 \end{figure}

We conducted a proton irradiation experiment for the PDD XRPIX, XRPIX6E, at the Heavy Ion Medical Accelerator in Chiba (HIMAC) in the National Institute of Radiological Sciences. 
The cross-sectional view of XRPIX6E is shown in Fig.~\ref{xr6e}.
XRPIX6E has $48~\times~48$ pixels and a pixel size of $36~{\rm \mu m} \times 36~{\rm \mu m}$~\cite{Harada19}.
The sensor layer has a thickness of $200~{\rm \mu m}$ and formed using a p-type floating zone wafer with a resistivity of $25~{\rm k\Omega \cdot cm}$.
The XRPIX is a back-illuminated sensor, and a p+ layer with a thickness of $0.2~{\rm \mu m}$ as an X-ray entrance window. 
{A} thin electrode is formed on the back side of the device {to apply a back-bias voltage}.
As introduced in Sec. \ref{sec:intro}, a sufficiently high doped BPW is formed at the Si--SiO$_2$ interface.
By applying a negative voltage ($V_{\rm PDD}$) to this p-well, the Si--SiO$_2$ interface is pinned to the applied voltage.

At HIMAC, we used two beamlines with 6-MeV and 100-MeV proton beams.
Fig.~\ref{p_expe6} shows the experimental setup with the 6-MeV proton beam.
The proton beam is scattered by a thin gold film with a thickness of 2.5 $\mu$m.
As shown in Fig.~\ref{p_expe6}, {proton beam scattered with an angle of 45 degrees} is irradiated on XRPIX6E, which is installed in {a vacuum chamber downstream of the gold film}. 
{A Faraday cup is installed to measure the unscattered beam to monitor fluctuations in the beam intensity during irradiation.} 
{Before irradiation on XRPIX6E, we measured the beam intensity at its position using an avalanche photodiode, and obtained the relationship between the scattered beam intensity measured with the avalanche photodiode and the unscattered beam intensity from the Faraday cup.}
{We estimated the total dose of XRPIX6E considering the variation in the intensity during the irradiation.}
Fig.~\ref{p_expe100} shows the experimental setup with the 100-MeV proton beam.
In this setup, the protons are directly incident on XRPIX6E.
The beam intensity was {again measured with} an avalanche photodiode before irradiating XRPIX6E.

We irradiated proton beams on the front side {(circuit layer side; see Fig.~\ref{xr6e})} of XRPIX6E.
The energies of the incident proton beams were $6$ and $100~{\rm MeV}$, which are sufficiently high to penetrate the oxide layer of XRPIX6E.
{The size of the proton beams was larger than the size of XRPIX6E ($4.5\times 4.5{\rm ~mm^2}$); hence the beam was uniformly irradiated on the peripheral circuit (e.g., column amplifier) on XRPIX6E as well as the pixel circuit.}
XRPIX6E was irradiated intermittently with proton beams to total doses of {$6~ {\rm krad}$} (6-MeV proton irradiation) and {$40~ {\rm krad}$} (100-MeV proton irradiation) in the BOX layer, and its performance between consecutive irradiation was evaluated.

XRPIX6E was cooled to {approximately $-65^{\circ} {\rm C}$} in the vacuum vessel during the proton beam irradiation. 
{A back-bias voltage of $-210~{\rm V}$ was applied for full depletion.}
The potential of the BPW was fixed by applying a voltage of $-2.0~{\rm V}$.

\section{Results of Irradiation Experiment }
\label{sec:RESEXP}

In this section, we present results of the proton irradiation experiment for the PDD XRPIX.
As an indicator of the performance degradation of the PDD XRPIX due to radiation damage, we measured its dark current, readout noise, chip output gain, and energy resolution while increasing the total dose applied by the proton beams.

\subsection{Dark current and Readout Noise}
\label{subsec:leak_ron}

\begin{figure}[tb]  
       \begin{center}
        \includegraphics[width=16cm]{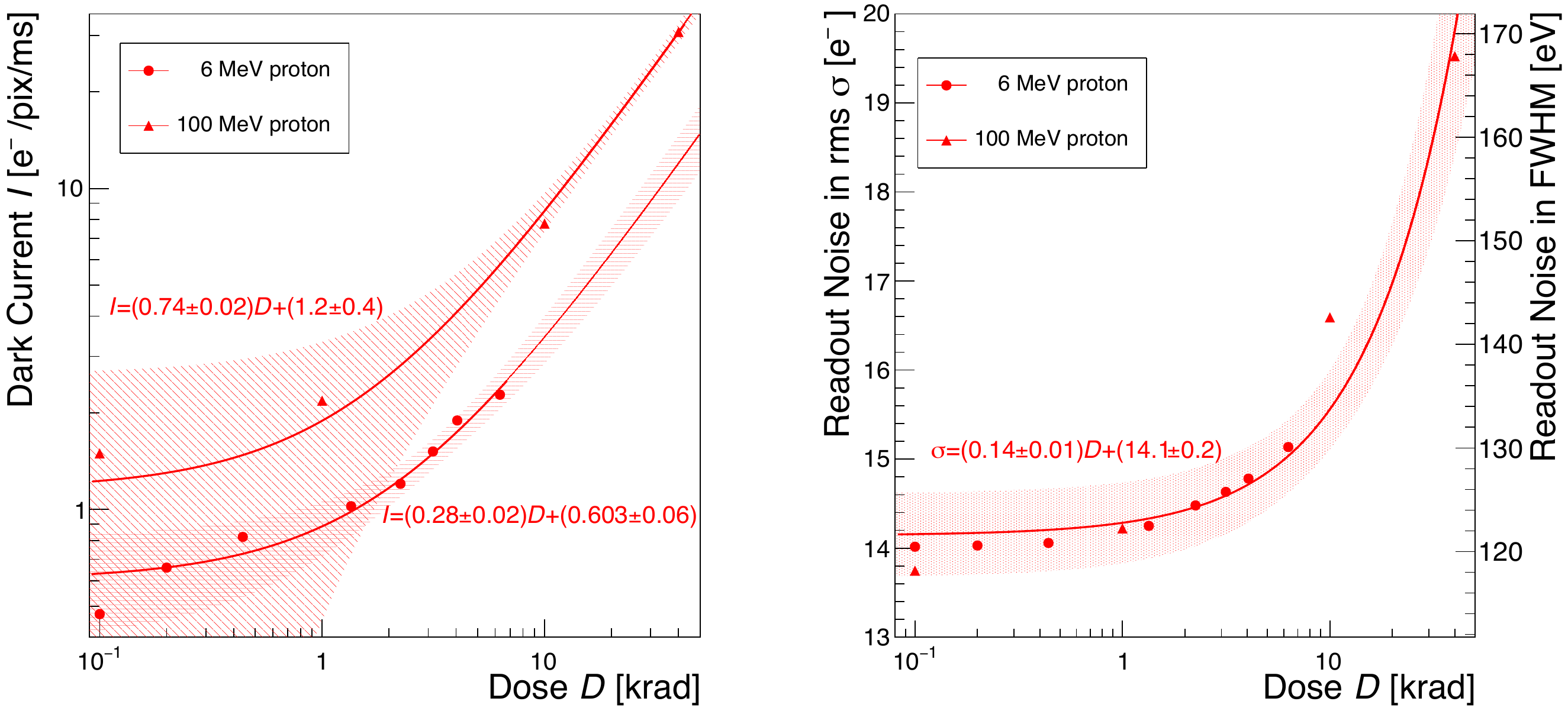}
 \end{center}
  \caption[Dark current and readout noise of PDD XRPIX as functions of total dose]{Dark current (left panel) and readout noise (right panel) of PDD XRPIX as functions of total dose. Solid lines and shaded regions represent best fit linear functions and {95\%} confidence regions, respectively. Data {points} at 0.1 krad {represent the} value before proton irradiation.}
\label{leakron}
\end{figure}

The dark current was {estimated} from the pedestal values.
{Because charge collected during an integration time is output as a signal, 
pedestal values increase with integration time due to dark current.}
{We measured the pedestal value as a function of the integration time and approximated the relationship between them as a linear function.}
{We then estimated the dark current from the slope of the function.}
The left panel {of} Fig.~\ref{leakron} shows dark currents of PDD XRPIX {as functions of total dose} for 6-MeV and 100-MeV irradiations. {Approximated linear functions are also plotted.}
Although there is a factor-of-two difference between 6-MeV and 100-MeV irradiations probably due to the individual characteristics of the device or experimental environment, the dark current is less than 30~$e^-$/pixel/ms even after the {$40$-krad} irradiation. Since the nominal integration time is {$1{\rm ~ms}$}, contribution of the dark current to the readout noise stays at {a low level of $\sqrt{30}~e^-\simeq 5\textrm{--}6~e^-$, which corresponds to the $\sim 10$\% increase in the readout noise}.

The readout noise was also evaluated based on the pedestal values with the nominal integration time of $1{\rm ~ms}$. 
We fitted {pedestal peaks of all pixels with} a single Gaussian function, and its standard deviation was used as a measure of the readout noise.
The right panel in Fig.~\ref{leakron} shows the degradation of the readout noise and a best-fit linear function. In this figure, both {the} 6-MeV and 100-MeV results are simultaneously fitted with a single linear function.
With a few kilorad irradiation required for the FORCE satellite, the readout noise increases by less than 10\%, staying below $15~e^-$ ({$130{\rm ~eV}$} in FWHM).
{However, it is larger than the required readout noise of $10~e^-$~\cite{Tsuru18}. Since the requirement is not satisfied even before the irradiation, it is necessary to reduce the readout noise in the next series of XRPIX.}
After the {$40$-krad} irradiation, the readout noise increases by $\sim 40\%$, up to {$20~e^-$ ($170{\rm ~eV}$} in FWHM).
Since the {contribution of dark current is approximately 10\%}, this degradation should be due to different noise components. We speculate that this is caused by an increase {in the} parasitic capacitance, and {we plans to conduct a comprehensive} study on this {issue} with device simulations.

\subsection{Noisy Pixel}
\label{subsec:noisy_pix}

\begin{figure}[tb]
        \centering
        \includegraphics[width=16cm]{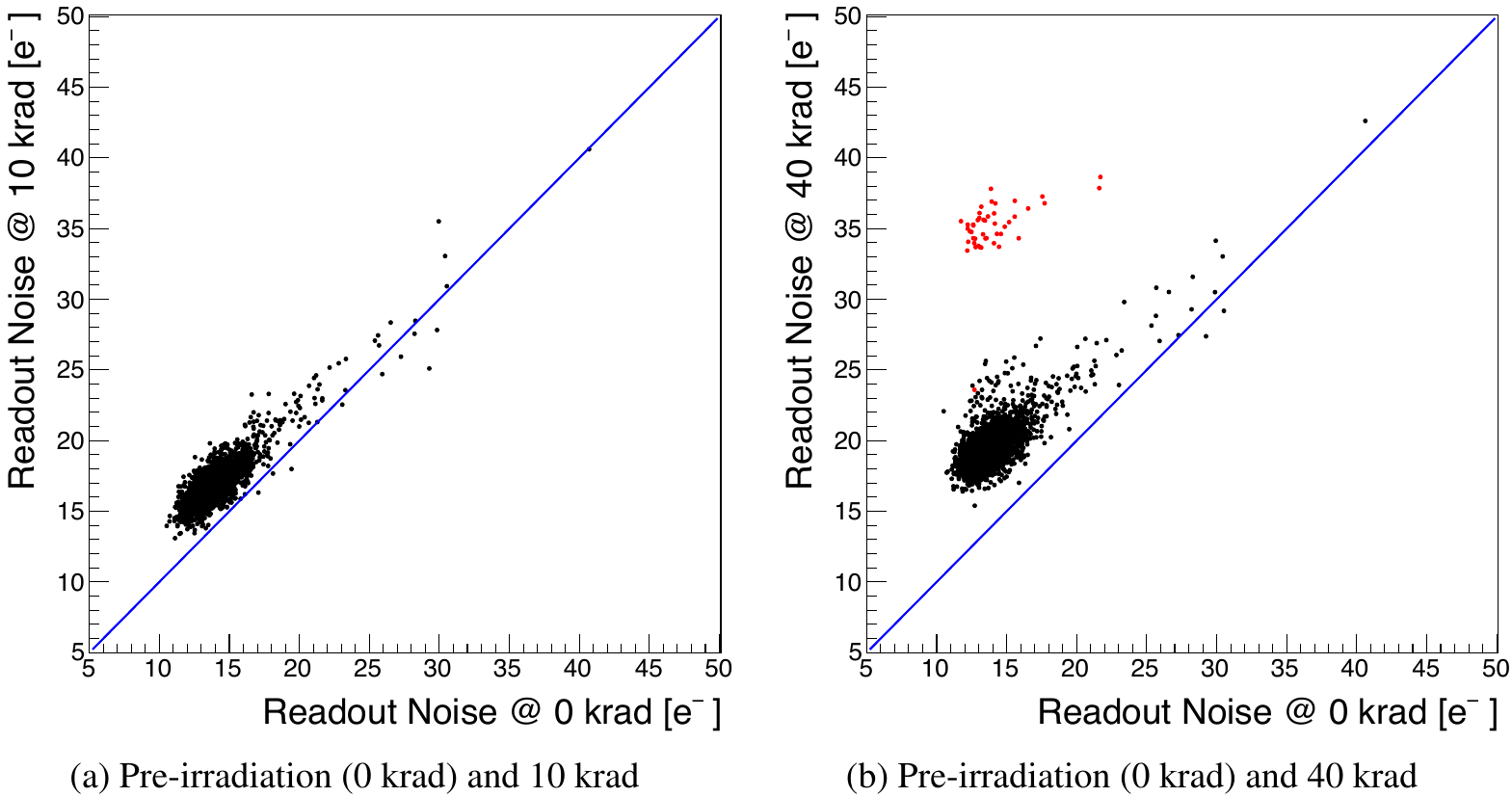}
        \label{fig:ron_cor}
    \vspace{2mm}
    \caption[Correlation between readout noises of each pixel of the PDD XRPIX before and after irradiation]{Correlation between readout noises of each pixel of the PDD XRPIX before and after irradiation. Solid blue line represents $y=x$ where the noise level of pre-irradiation equals to that of post-irradiation. Red dots show results for {the column address} ${\rm ca}=30$, where readout noise is particularly high after 40-krad irradiation.} 
    \label{fig:ron_cor}
\end{figure}

\begin{figure}[tb]
        \centering
        \includegraphics[width=16cm]{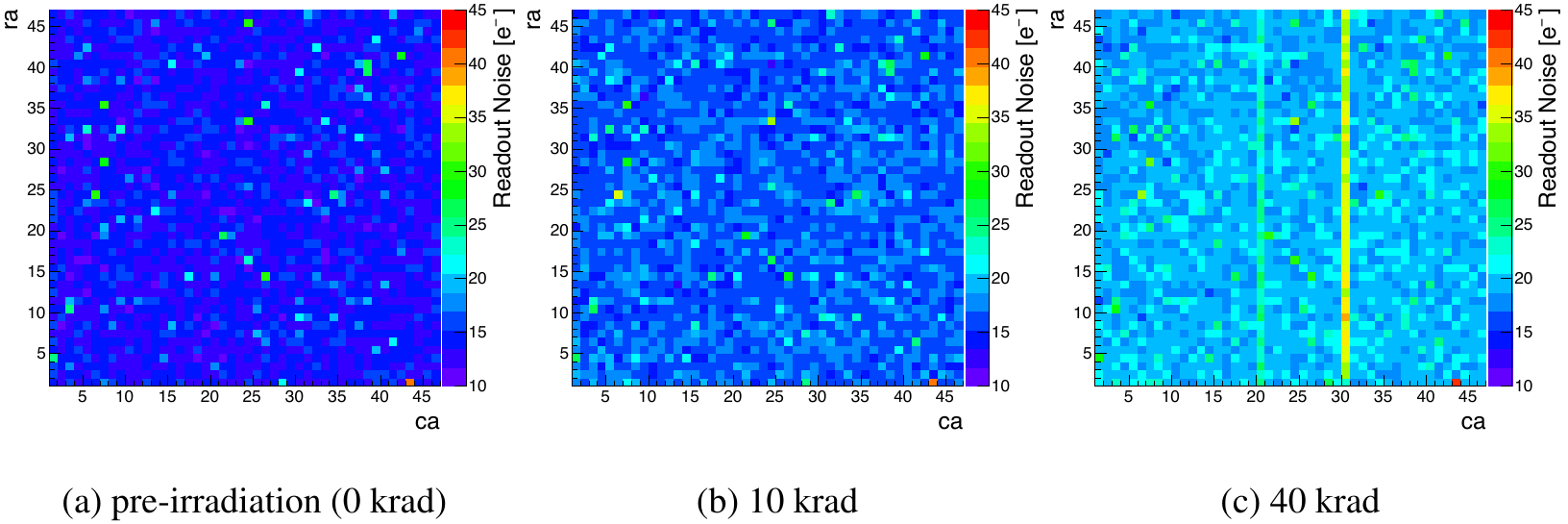}
    \vspace{2mm}
    \caption[Two-dimensional map of readout noises of each pixel of the PDD XRPIX]{Two-dimensional map of readout noises of each pixel of the PDD XRPIX (a)before irradiation, (b)after {$10$-krad} irradiation, and (c){$40$-krad} irradiation.}
    \label{fig:ron_map}
\end{figure}

\begin{figure}[tb]
    \begin{minipage}[b]{0.49\hsize}
        \centering
        \includegraphics[width=8cm]{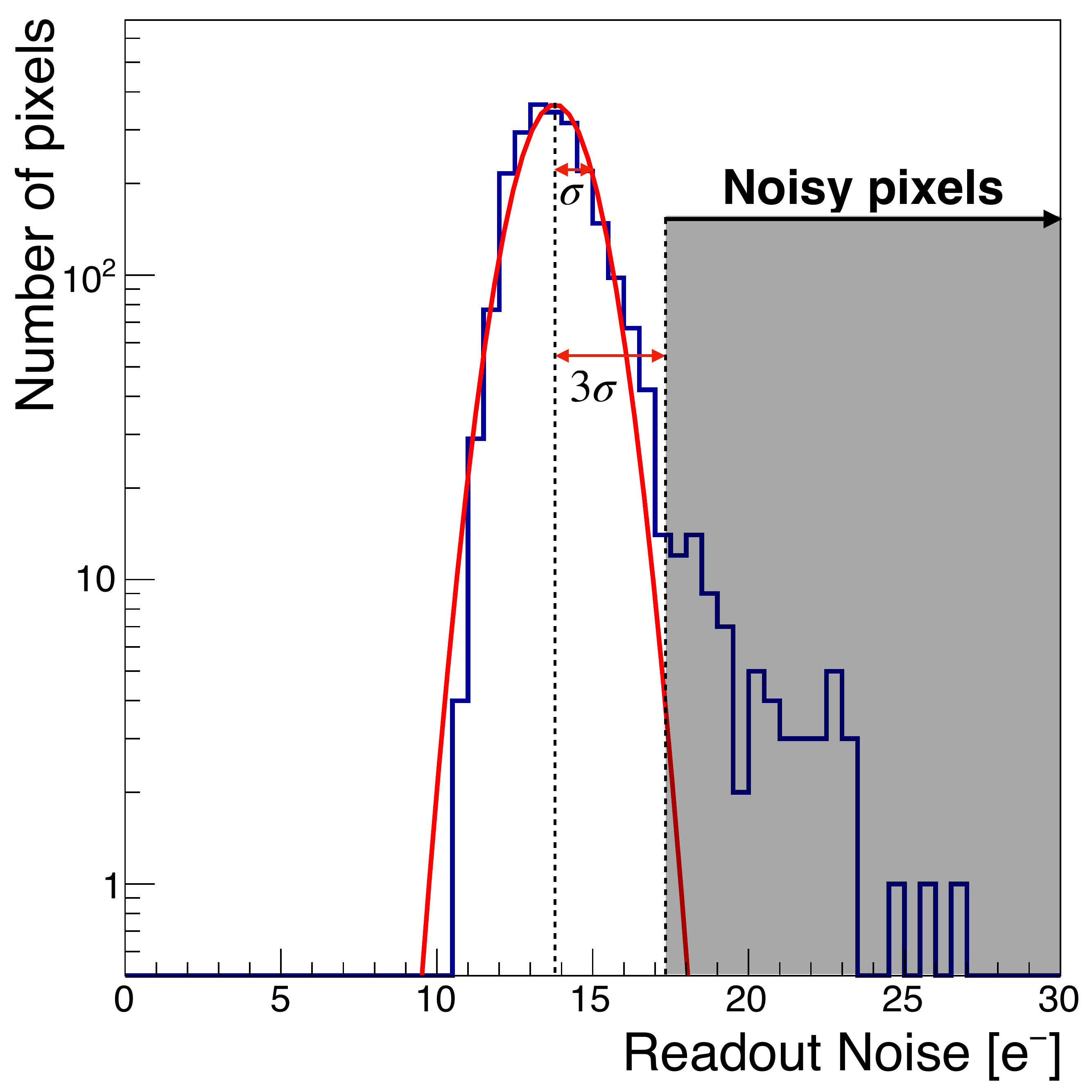}
        \caption[Histogram of readout noises for all pixels before irradiation]{Histogram of readout noises for all pixels before irradiation. Definition of noisy pixels is shown by shading.}
        \label{fig:sigma_hist}
    \end{minipage}
    \hspace{2mm}
    \begin{minipage}[b]{0.49\hsize}
        \centering
        \includegraphics[width=8cm]{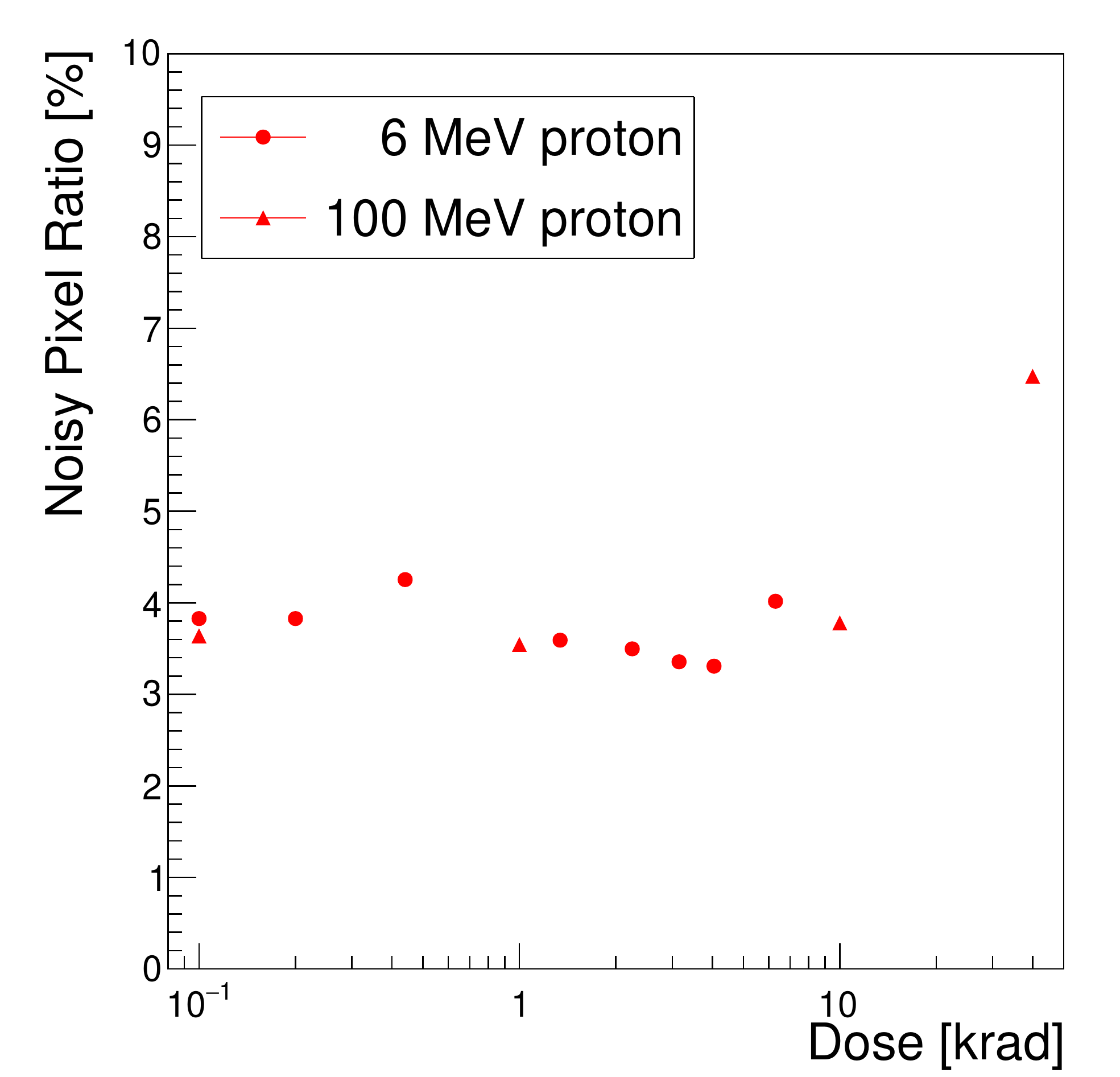}
        \caption[Noisy pixel ratio of PDD XRPIX as function of total dose]{Noisy pixel ratio of PDD XRPIX as function of total dose.\\}
        \label{fig:badpix}
    \end{minipage}
\end{figure}

We further investigated the readout noise of each pixel.
Fig.~\ref{fig:ron_cor} shows the correlation between the readout noises of each pixel before and after irradiation.
As shown in Fig.~\ref{fig:ron_cor}(a), most data points shift uniformly upward.
Furthermore, as can be seen {in} Fig.~\ref{fig:ron_cor}(b), after the 40-krad irradiation, {in addition to} the overall noise level increase{, there is an emergence of} some pixels with more significant increase in the noise than the other pixels appear.

Fig.~\ref{fig:ron_map} shows two-dimensional maps of the readout noises of each pixel before and after irradiation.
The increasing trend of the noise is uniform for all pixels.
However, after the 40-krad irradiation, the noise levels of the pixels at column addresses ${\rm ca}=20$ and ${\rm ca}=30$ are higher than those of the pixels in the other columns.
The dots shown in red in Fig.~\ref{fig:ron_cor}(b) are the results for the pixels with ${\rm ca}=30$, where the readout noises are particularly high.
This column-dependent feature is {attributed} to an increase in the noise originating from the circuitry after the column amplifier.

The number of ``noisy pixels'' is also an important indicator of the performance degradation.
A noisy pixel is defined using the distribution of the readout noises of each pixel, as shown in Fig.~\ref{fig:sigma_hist}.
The mean and standard deviation, $\sigma$, of the distribution were estimated by fitting the distribution with a Gaussian function.
Subsequently, we defined a noisy pixel as a pixel with readout noise above $3\sigma$ from the mean value.
{Fig.~\ref{fig:badpix} presents} a ratio of the number of noisy pixels to the total number of pixels as a function of the total dose. 
The noisy pixel ratio of the PDD XRPIX is {approximately} 4\% before the irradiation, and suddenly increases after the 40-krad irradiation.
The reason for this increase is that the pixels with ${\rm ca}=30$ have higher readout noises than the other pixels, as mentioned above. Since there are 48 columns in XRPIX6E, the sudden increase by $1/48\simeq 2\%$ after the 40-krad irradiation is consistent with the number of pixels with ${\rm ca}=30$.
{Regarding} the spectral performance, because noisy pixels are less than $10\%$ even after the 40-krad irradiation, their contribution to the evaluation of energy resolution using all pixels is negligible.

\subsection{X-ray Spectral Performance}
\label{subsec:spec_per}
\begin{figure}[tb]  
     \begin{center}
        \includegraphics[width=15cm]{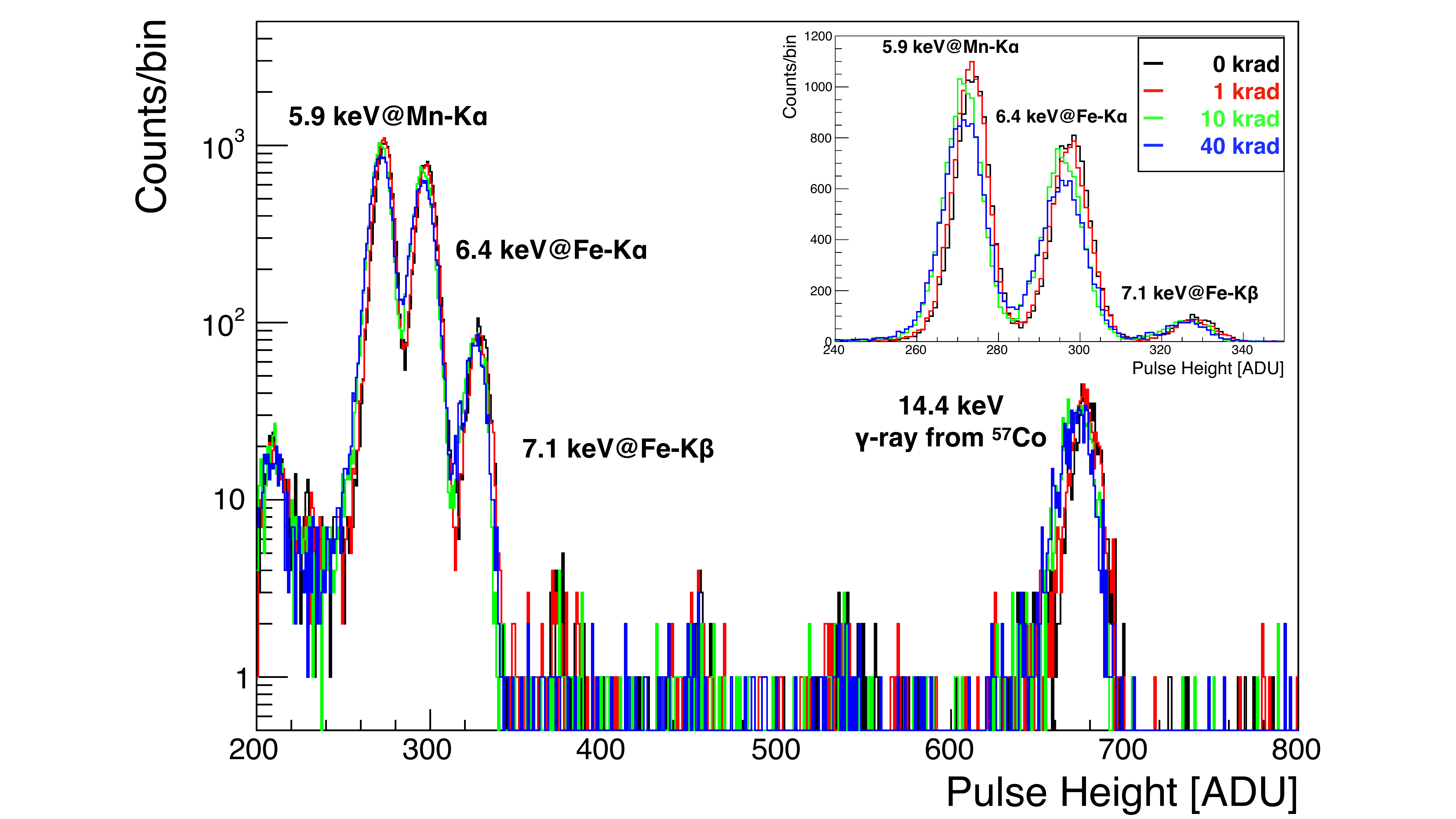} 
 \caption[X-ray spectra of the PDD XRPIX after proton irradiation from $^{55}$Fe and $^{57}$Co]{X-ray spectra of the PDD XRPIX after proton irradiation from $^{55}$Fe and $^{57}$Co. Small panel on right side of figure is magnified view of spectrum in energy range of approximately 5--8 keV.}
 \label{spec}
 \end{center}
 \end{figure}

The spectral performance was evaluated by irradiating X-rays from $^{55}$Fe and $^{57}$Co radioisotopes.
Fig.~\ref{spec} shows the corresponding X-ray spectra after the proton irradiation.
The spectral shape after the $1$-krad irradiation was similar to that before the irradiation; however, the peak position of the energy spectrum was found to shift slightly toward lower pulse height with the increase in the total irradiation dose.

\begin{figure}[tb]  
       \begin{center}
        \includegraphics[width=16cm]{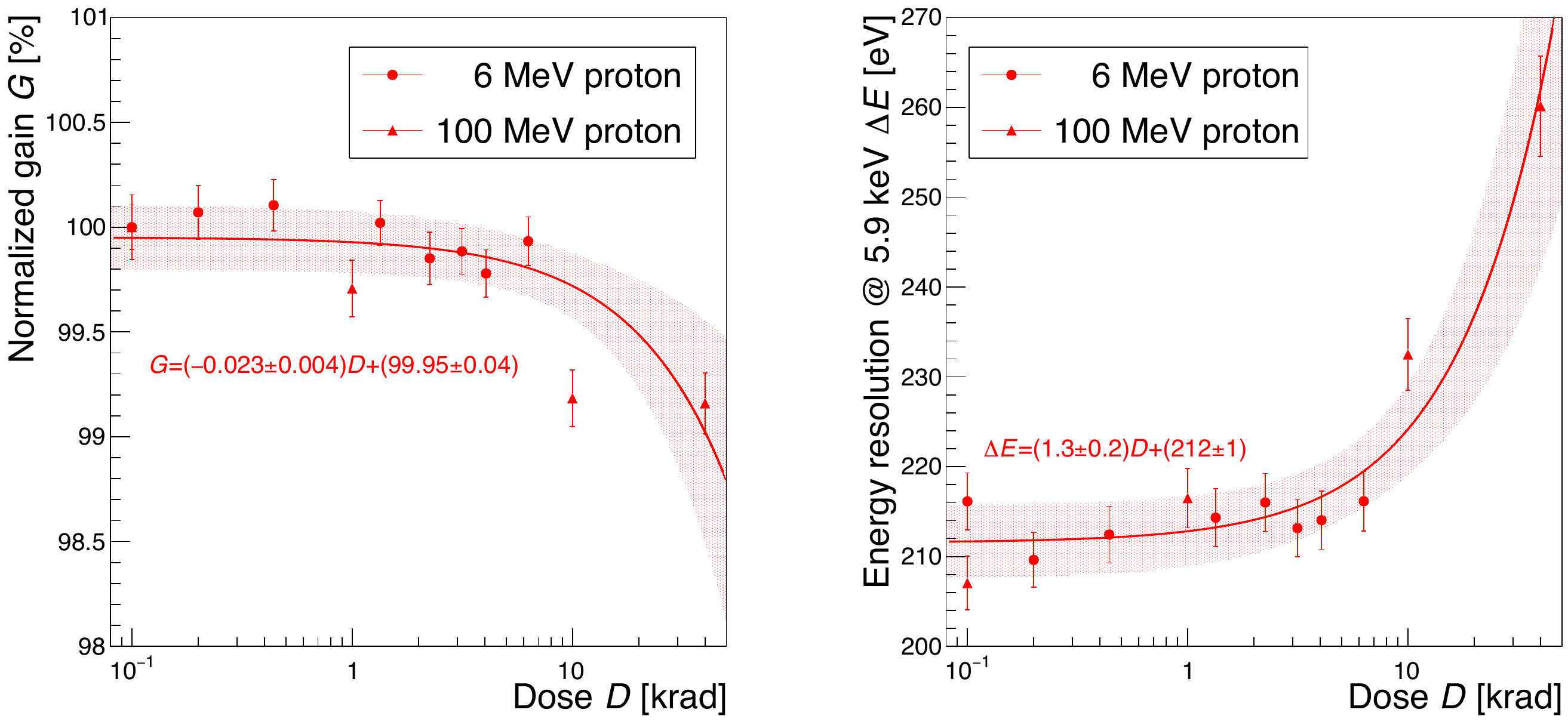}
 \end{center}
  \caption[Gain and energy resolution of PDD XRPIX as functions of total dose]{Gain (left panel) and energy resolution (right panel) of PDD XRPIX as functions of total dose. Best fit linear functions and {95\%} confidence regions are overplotted similarly as in Fig.~\ref{leakron}.}
\label{gainfwhm}
 \end{figure}

For a quantitative evaluation, we fitted the emission lines of ${\rm Mn~K\alpha}$ at $5.9~{\rm keV}$, ${\rm Fe~K\alpha}$ at $6.4~{\rm keV}$, and ${\rm Fe~K\beta}$ at $7.1~{\rm keV}$, and $\gamma$-rays from $^{57}$Co at $14.4~{\rm keV}$, with Gaussian {functions}.
To obtain the chip gain, we fitted the best-fit peak positions with a linear function of X-ray/$\gamma$-ray energies (i.e., energy calibration curves) and extracted the gain from the slope of the function.
The left panel of Fig.~\ref{gainfwhm} shows the gain as a function of the total dose. The best-fit linear function of the simultaneous fitting of both 6-MeV and 100-MeV results are also shown in the figure.
Because the gain of the PDD XRPIX used for the 6-MeV irradiation ($47.41\pm0.04~{\rm \mu V/e^-}$ at 0~rad) {differs} from that for {the} 100-MeV irradiation ($47.09\pm0.05~{\rm \mu V/e^-}$ at 0~rad), in the figure, the gain is normalized using the pre-irradiation value.
In the range of a few {kilorads}, the gain degradation is less than $0.5\%$.
After the {$40$-krad} irradiation, the gain {decreased} by $0.84 \pm 0.14 \% $ compared to that before the irradiation.

The energy resolution was evaluated from the FWHM of the optimal Gaussian function for $5.9~{\rm keV}$.
The right panel of Fig.~\ref{gainfwhm} shows the energy resolution as a function of the total dose.
The best-fit linear function of the simultaneous fitting of both 6-MeV and 100-MeV results are also shown in the figure.
With a few kilorad irradiation, the energy resolution remains constant within a few {percentages}.
After the {$40$-krad} irradiation, the energy resolution is degraded by $25 \pm 3 \% $ compared to that before the irradiation.
However, the energy resolution even after the {$40$-krad} irradiation, equivalent to $\sim400$~years in the FORCE orbit, is better than the required performance of FORCE, i.e., 300 eV at 6 keV.
Therefore, {we conclude} that the PDD XRPIX has sufficient radiation hardness.


\section{Discussion}

\subsection{Comparison with DSOI}
\begin{figure}[tb]  
     \begin{center}
        \includegraphics[width=15cm]{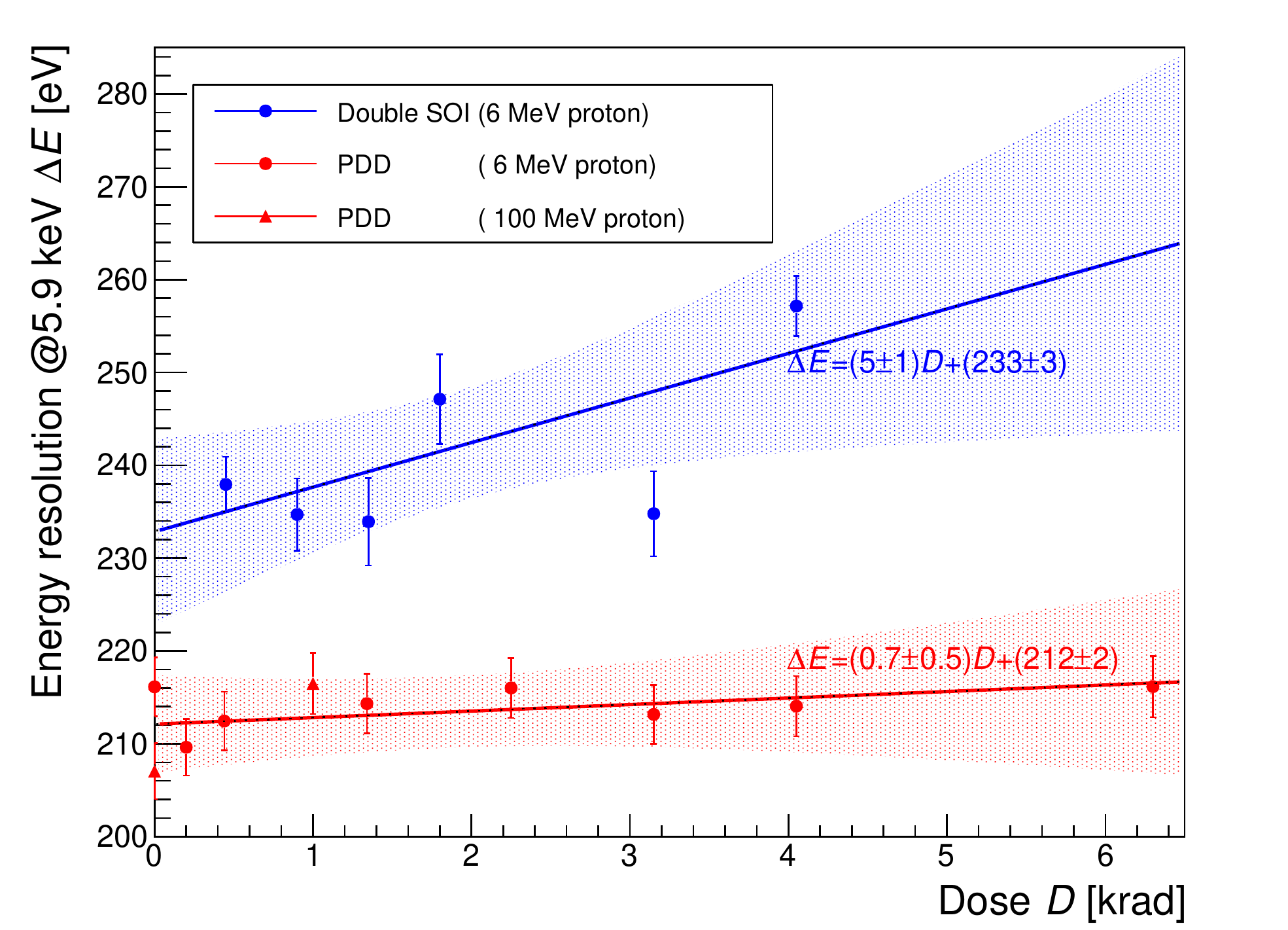} 
 \caption[Energy resolution of PDD XRPIX and DSOI XRPIX as a function of total dose below $6$~krad]{Energy resolution of PDD XRPIX (red) and DSOI XRPIX (blue) as a function of total dose below {$6$~krad}. Best fit linear functions and {95\%} confidence regions are overplotted similarly as in Fig.~\ref{leakron}.}
 \label{fwhm}
 \end{center}
 \end{figure}

Fig.~\ref{fwhm} shows a comparison of the energy resolutions between PDD XRPIX and the previous XRPIX, DSOI XRPIX, which was introduced in Sec.~\ref{sec:intro}. 
{For a comparison under} the same condition, {data from PDD XRPIX are plotted only} below 6.5~krad, which is the maximum total dose of the proton irradiation experiment of DSOI XRPIX~\cite{Hagino21}. Both results of PDD and DSOI are fitted with linear functions in this dose range.
According to the fitting, after the {$4$-krad} irradiation, the energy resolution of the DSOI device is degraded by $7.0 \pm 2.5~\%$, whereas that of the PDD device remains constant within $1\%$.
{Therefore}, the radiation hardness is improved in the PDD XRPIX, at least in the aspect of energy resolution.

One possible reason for this improvement is related to structural differences near the Si--SiO$_2$ interface between PDD and DSOI.
The problems with the DSOI XRPIX are dark current and charge loss at the Si--SiO$_2$ interface.
Because, in the DSOI structure, a significant fraction of the interface region is exposed to the depletion region in the sensor layer, electrons are generated through the interface state, resulting in a large dark current.
Furthermore, because signal charges are trapped at the interface during charge collection, charge loss occurs, and the energy resolution deteriorates~\cite{Hagino19}.

In the PDD structure, {on the other hand,} the interface is covered with the BPW, which is a sufficiently high doped p-well; therefore, the dark current is reduced by the recombination of the electrons and holes generated by the BPW.
In addition, the charge loss can be suppressed by changing the electric field structure with the bias voltage {applied to} the BPW, ensuring {that} the signal charges are collected without passing near the interface.
Consequently, the charge collection efficiency is improved, and the energy resolution is enhanced~\cite{Harada19,Kayama19}.

{Regarding the} radiation damage, the interface state increase {owing to} the radiation probably contributes to the degradation of the energy resolution of the DSOI XRPIX. Consequently, in the PDD XRPIX, because the Si--SiO$_2$ interface is covered by the BPW, this effect is suppressed, as explained above. Therefore, one of the reasons for the improvement in the radiation hardness of the PDD XRPIX is that the BPW reduces the effect of the increased interface states caused by the radiation.

\subsection{TID Compensation with Fixed-Potential Layer}
 
 \begin{figure}[tb]  
 \begin{tabular}{cc}
\begin{minipage}{0.48\hsize}
        \begin{center}
        \includegraphics[width=8cm]{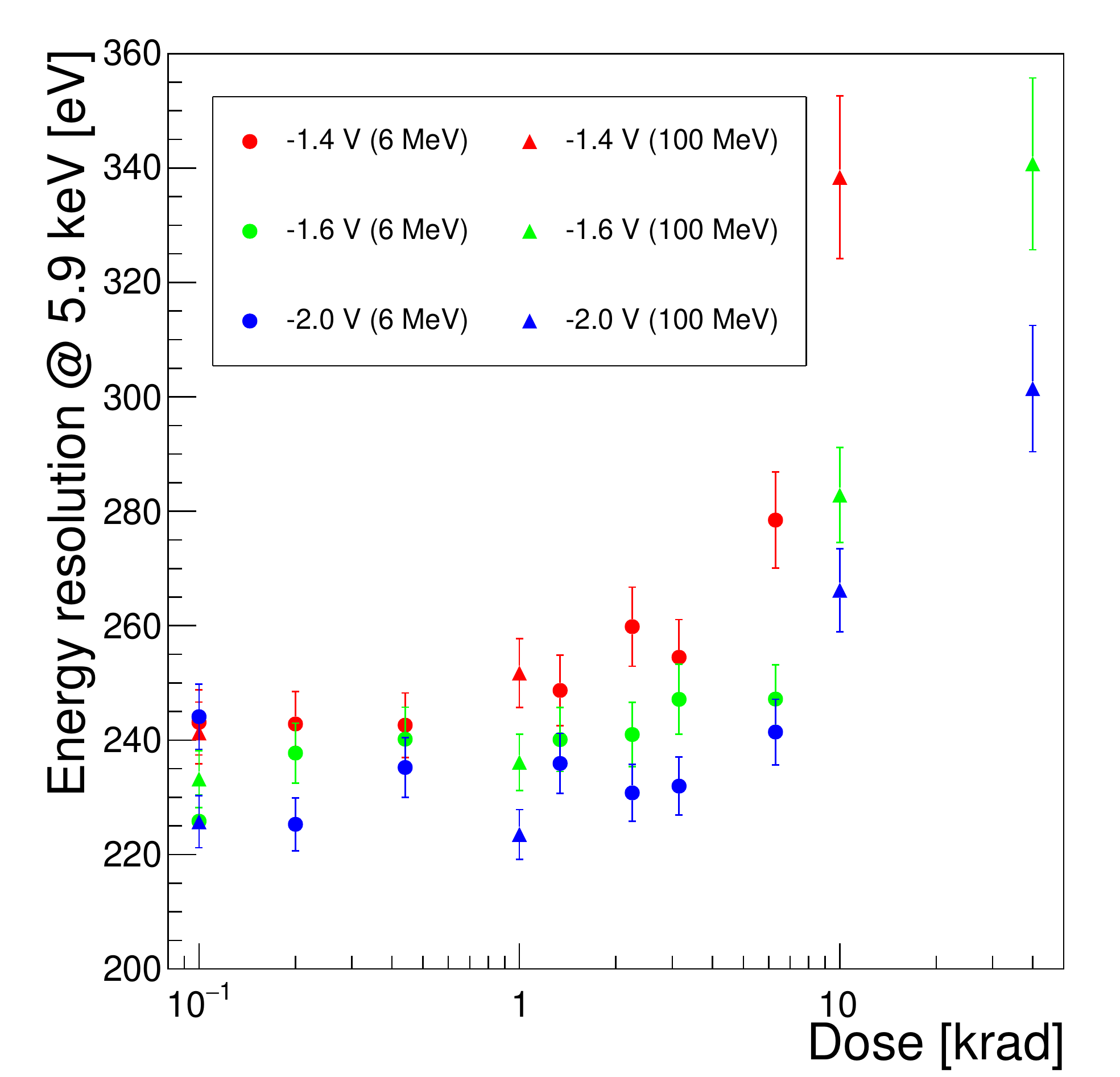} 
        \caption[Relation between energy resolution and total dose when voltage applied to fixed-potential layer is changed]{Relation between energy resolution and total dose when voltage applied to fixed-potential layer is changed to $-1.4~{\rm V}$ (red color), $-1.6~{\rm V}$ (light green color), and $-2.0~{\rm V}$ (blue color). Data point at 0.1 krad represents {the} value before proton irradiation.}
        \label{vpdd}
 \end{center}
\end{minipage}

\hspace{2mm}

\begin{minipage}{0.48\hsize}
      \begin{center}
        \includegraphics[width=8cm]{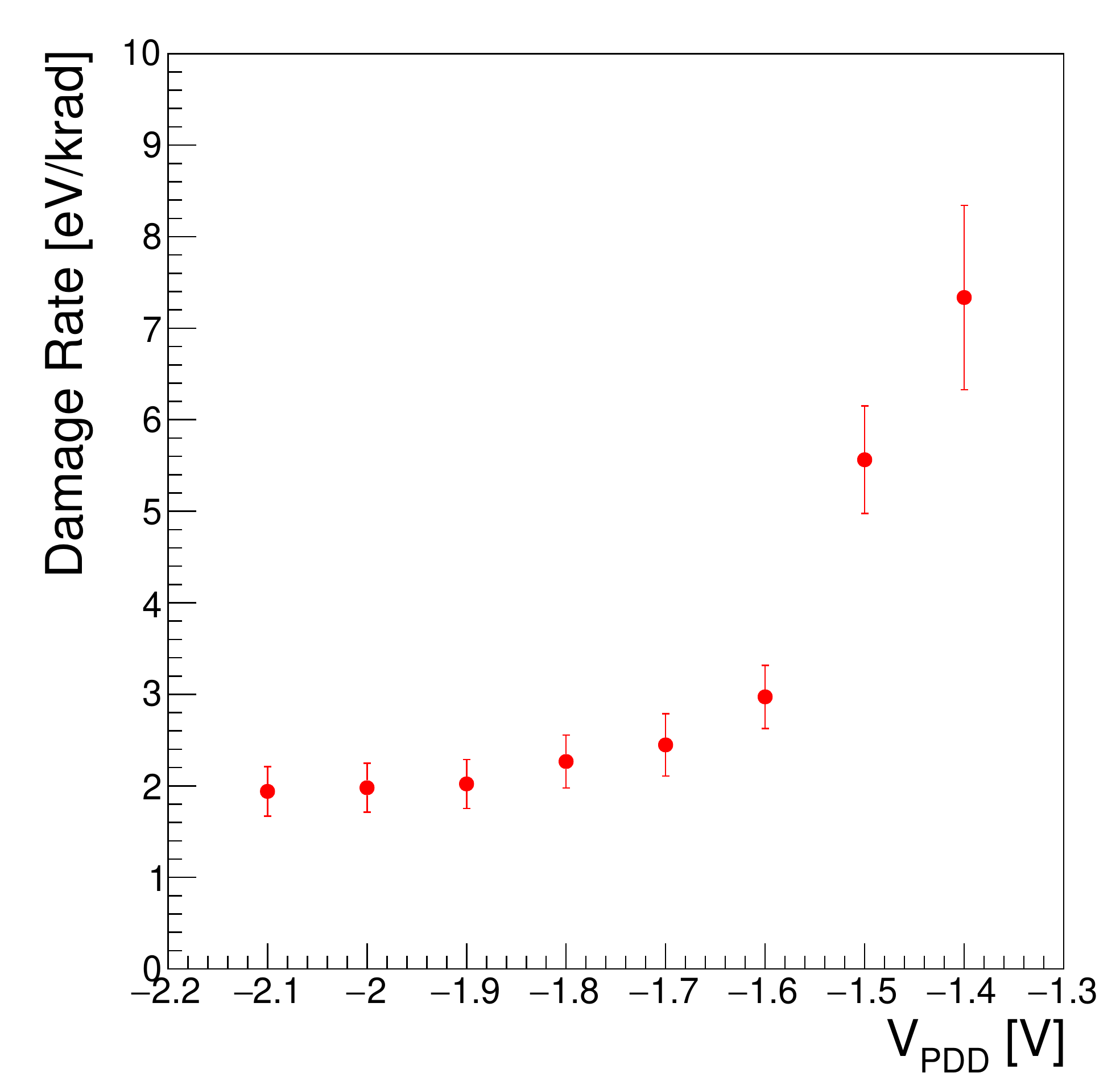} 
        \caption[Relation between damage rate and voltage applied to fixed-potential layer]{Relation between damage rate and voltage applied to fixed-potential layer. Application of high negative voltage is associated with significant suppression of degradation of energy resolution.\\}
        \label{vpdd_gra}
 \end{center}
\end{minipage}
\end{tabular}
 \end{figure}

To investigate the compensation of the TID effect by the PDD structure, we evaluated the energy resolution by changing the {bias} voltage applied to the BPW.
The bias fixes the potential at the Si--SiO$_2$ interface, forming a fixed-potential layer.
It is considered that the negative potential of the Si--SiO$_2$ interface cancels out the effect of the holes accumulated in the oxide layer (TID compensation).
Figs.~\ref{vpdd} and ~\ref{vpdd_gra} show the energy resolution of XRPIX6E when the applied voltage is changed.
We changed the voltage from $-1.4~{\rm V}$ to $-2.1~{\rm V}$ in $0.1~{\rm V}$ steps.
{The energy resolution stays almost constant independently of the applied voltage below a few kilorads.}
However, as the total dose increases, a {smaller} absolute value of the applied voltage {results in more} significant degradation of the energy resolution. 
{A high negative voltage can help to suppress the degradation of the energy resolution.}

To more clearly show the dependence on the negative voltage applied to the fixed potential layer, $V_{\rm PDD}$, we fitted the energy resolutions at each $V_{\rm PDD}$ with a linear function of {a} total dose. Since the slope of the best-fit function represents the speed of degradation, we refer to it as damage rate, and plotted it as a function of total dose in Fig.~\ref{vpdd_gra}.
It can be clearly seen that the change with the total dose depends on the negative voltage applied to the fixed-potential layer.
With a small absolute value of $V_{\rm PDD}$, the degradation of the energy resolution is enhanced.
The effect of the accumulated holes {seems to be} suppressed by applying a negative voltage to the fixed-potential layer, i.e., the TID effect is suppressed.

\subsection{Causes of Degradation of Spectral Performance}
The proton irradiation experiment of XRPIX6E revealed the degradation of the spectral performance.
After the {$40$-krad} irradiation, the gain and energy resolution were degraded by $0.84 \pm 0.14 \% $ and $25 \pm 3 \% $, respectively, compared to {those} before irradiation. In this section, we discuss possible causes of these degradations.

\subsubsection{Gain}
The gain degradation is considered to be caused by the decrease in the gain of the charge sensitive amplifier due to the increase in the parasitic capacitance.
In a previous study on the DSOI XRPIX, the device simulation revealed that the size of the buried n-well around the sense node increases owing to the oxide-trapped positive charges, thereby increasing the parasitic capacitance.
This increased parasitic capacitance is quantitatively consistent with that expected from the observed gain decrease of the DSOI XRPIX~\cite{Hagino21}.

According to the previous study on the DSOI XRPIX~\cite{Hagino21}, the inverse of the gain $G$ changes with the parasitic capacitance of the sense node $C_{\rm SN}$ as
\begin{equation}
\Delta\left(\frac{1}{G}\right)=\frac{1}{AG_{\rm SF}G_{\rm SH}}\Delta C_{\rm SN}\simeq 1.4\times 10^{-2} \left(\frac{\Delta C_{\rm SN}}{1~{\rm fF}}\right)~{\rm fF},
\end{equation}
where $A\simeq108$, $G_{\rm SF}\simeq0.82$, and $G_{\rm SH}\simeq0.8$ {represent} the open-loop gain of the {charge sensitive amplifier}, source follower circuit gain, and gain from sample-hold to the output buffer circuit, respectively~\cite{Takeda20}. Since the gain of PDD XRPIX was degraded from $47.1{\rm ~\mu V/e^{-}}$ at 0~rad to $46.7{\rm ~\mu V/e^{-}}$ at 40~krad, the change in the inverse of the gain is $\Delta\left( 1/G\right)\simeq 2.9\times10^{-2}{\rm ~fF}$. {Therefore}, if the parasitic capacitance increases by {$2{\rm ~fF}$} with the 40-krad irradiation, the experimental result can be explained by this mechanism.
Although the existence of such a parasitic capacitance is not unreasonable, to fully resolve the gain degradation mechanism of PDD XRPIX, we should investigate with the device simulation, which is our future work.

\subsubsection{Energy resolution}
To investigate the degradation of the energy resolution, we evaluated its energy dependence.
The energy dependence of the energy resolution is presumed to be described as follows:
 
\begin{equation}
\label{eq:delEE}
\Delta E=2.355W\sqrt{\left(\frac{FE}{W} \right)+\sigma_{\rm inde}^2+\left(\frac{\sigma_{\rm gain}E}{W} \right)^2}\ ,
\end{equation}
where $\Delta E$, $E$, $F$, $W$, $\sigma_{\rm inde}$, and $\sigma_{\rm gain}$ {represent} the energy resolution, X-ray/$\gamma$-ray energy, Fano factor (0.12), mean ionization energy per electron--hole pair in Si ($3.65~{\rm eV}$), energy-independent noise in rms, and pixel-to-pixel relative gain variation in rms, respectively~\cite{Kodama21}. 
As {expressed in} Eq. \ref{eq:delEE}, the Fano noise, energy-independent components (e.g., readout noise), and energy-dependent pixel-by-pixel pedestal nonuniformity contribute to the energy resolution.

\begin{figure}[tb]  
    \begin{tabular}{cc}
        \begin{minipage}{0.49\hsize}
            \begin{center}
                \includegraphics[width=8cm]{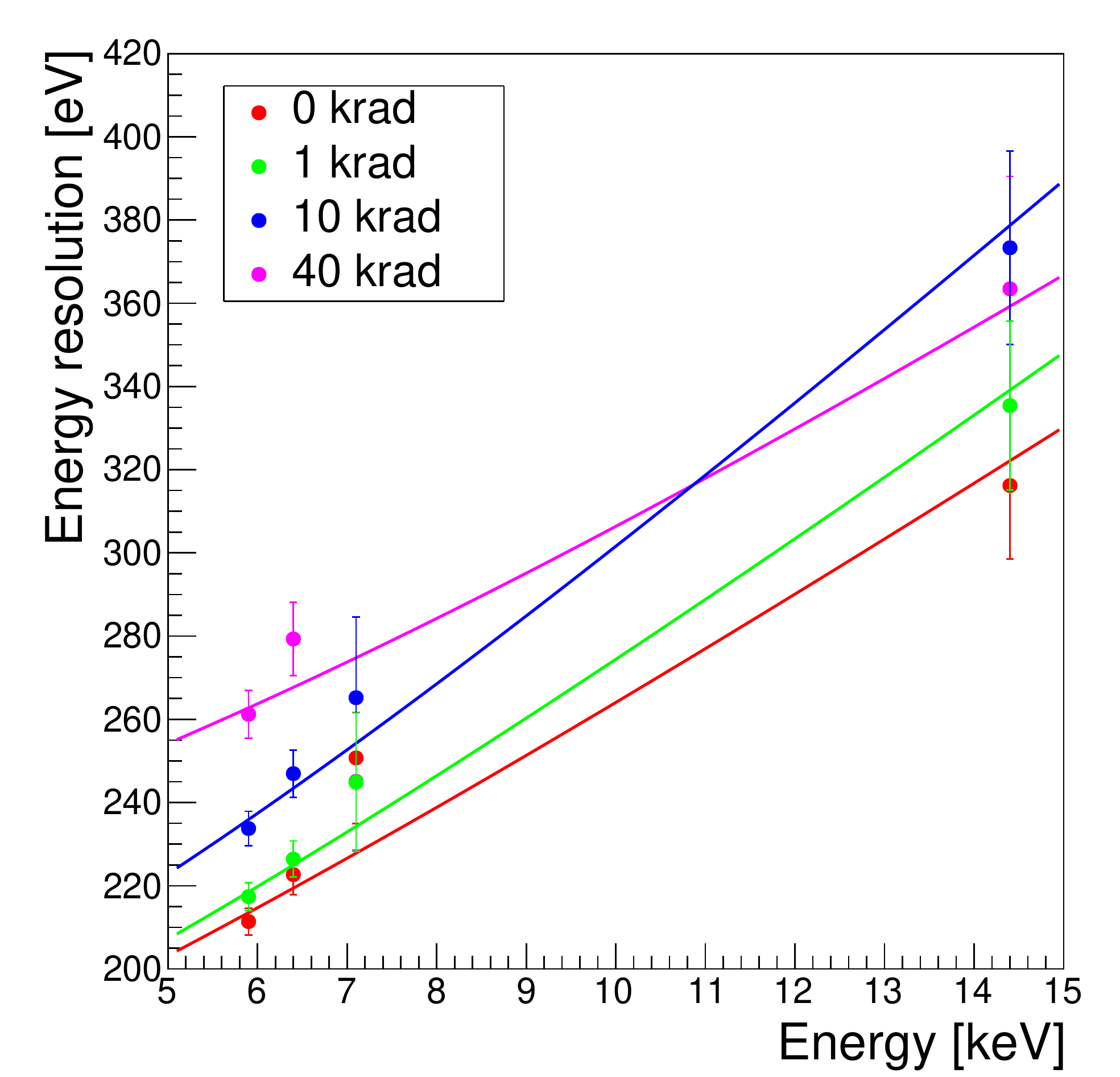} 
                \caption[Relation between energy resolution and X- and $\gamma$-ray energies]{Relation between energy resolution and X- and $\gamma$-ray energies at $0~{\rm krad}$ (red color), {$1~{\rm krad}$} (light green color), {$10~{\rm krad}$} (blue color), and {$40~{\rm krad}$} (magenta color). Solid lines represent best results of fitting using Eq.~\ref{eq:delEE}.}
                \label{gain_vari}
            \end{center}
        \end{minipage}
        
        \hspace{2mm}
        
        \begin{minipage}{0.48\hsize}
                \begin{center}
                \includegraphics[width=8cm]{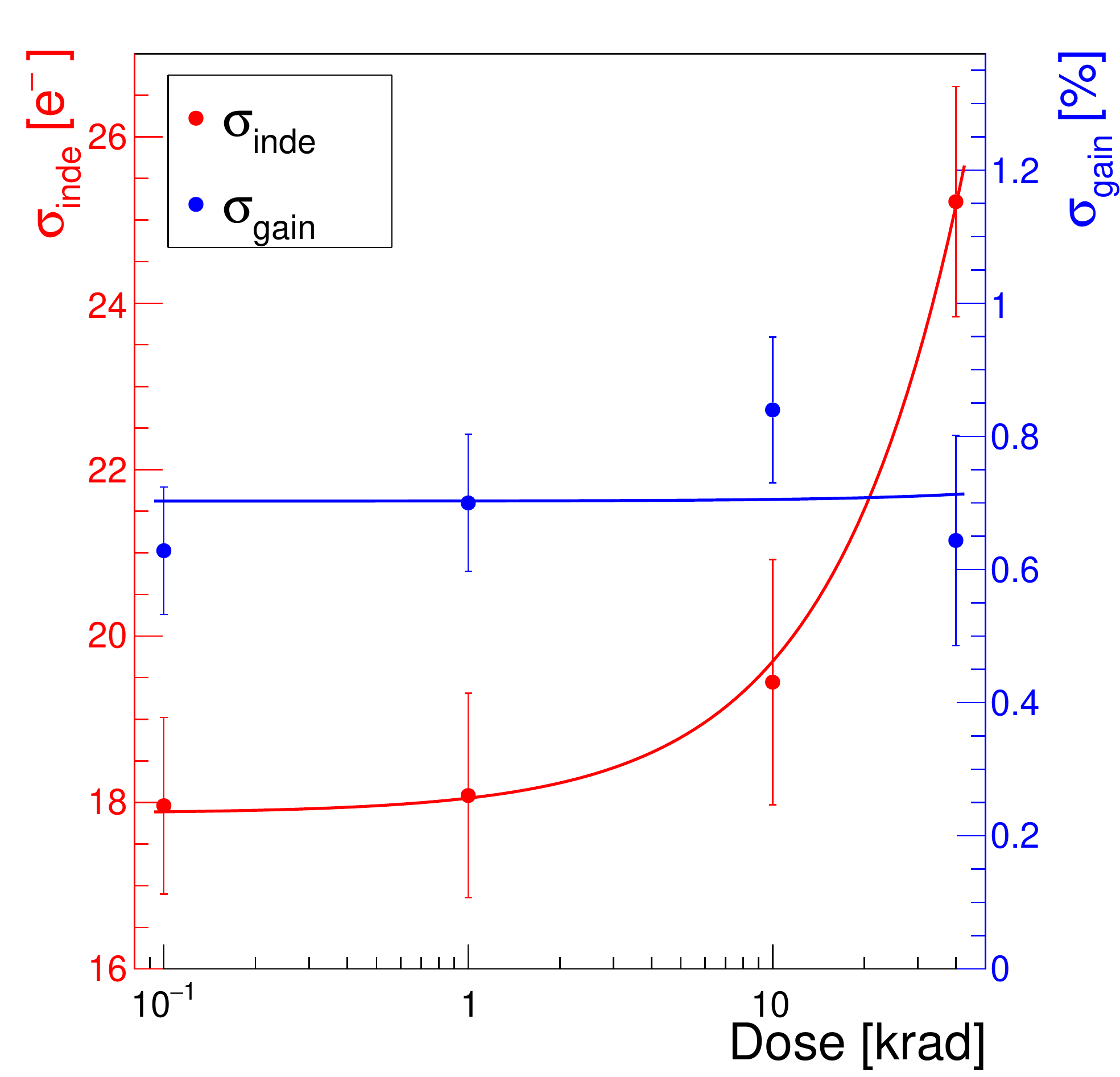} 
                \caption[$\sigma_{\rm inde}$ and $\sigma_{\rm gain}$ as functions of total dose]{$\sigma_{\rm inde}$ (red color) and $\sigma_{\rm gain}$ (blue color) as functions of total dose. Data point at 0.1 krad represents value before proton irradiation. Solid lines represent best fit linear functions.\\}
            \label{noise_comp}
            \end{center}
        \end{minipage}
    \end{tabular}
\end{figure}

%
%

Fig.~\ref{gain_vari} shows the relationship between each energy resolution and the X-ray and $\gamma$-ray energies.
To obtain $\sigma_{\rm inde}$ and $\sigma_{\rm gain}$, we fitted this relationship {with} Eq.~\ref{eq:delEE}.
Fig.~\ref{noise_comp} shows the fitting result at each total dose.
Although the gain changed after the {$40$-krad} irradiation, as described in Sec. \ref{subsec:spec_per}, $\sigma_{\rm gain}$ {did} not increased.
However, after the {$40$-krad} irradiation, $\sigma_{\rm inde}$ increased by $40\pm 11\%$.
Therefore, the degradation of the energy resolution would be due to increasing $\sigma_{\rm inde}${, and the contribution of the pixel-to-pixel gain variation $\sigma_{\rm gain}$ was not significantly changed}.
This conclusion is {supported also} by the result that the readout noise increased by $42.0\pm0.2\%$ after the {$40$-krad} irradiation (c.f., Sec. \ref{subsec:spec_per}).

\section{Conclusions}
We evaluated the radiation hardness of a new XRPIX with a PDD structure by irradiating 6- and 100-MeV proton beams at the HIMAC. 
We found that even after irradiation of {$40~{\rm krad}$}, equivalent to $\sim400$~years in the FORCE orbit, the energy resolution satisfied the required performance of FORCE ($300~{\rm eV}$ at $6~{\rm keV}$), and the PDD XRPIX {exhibited} better radiation hardness than the DSOI XRPIX.
It was also found that the TID effect was suppressed by applying a negative potential to the fixed-potential layer of the PDD structure.
In addition, we investigated the degradation of the energy resolution; it was determined that the degradation would be due to increasing energy-independent components, e.g., readout noise.

\acknowledgments 
We acknowledge the relevant advice and manufacture of the XRPIXs by the personnel of LAPIS Semiconductor Co., Ltd. This study was based on the results of the experiments conducted at Heavy Ions at NIRS-HIMAC. This study was supported by MEXT/JSPS KAKENHI Grant-in-Aid for Scientific Research on Innovative Areas 25109002 (Y.A.) and 25109004 (T.G.T., T.T., K.M., A.T., and T.K.), Grant-in-Aid for Scientific Research (B) 25287042 (T.K.), Grant- in-Aid for Young Scientists (B) 15K17648 (A.T.), Grant-in-Aid for Challenging Exploratory Research 26610047 (T.G.T.), and Grant-in-Aid for Early-Career Scientists 19K14742 (A.T.). This study was also supported by the VLSI Design and Education Center (VDEC), Japan and the University of Tokyo, Japan in collaboration with Cadence Design Systems, Inc., USA; Mentor Graphics, Inc., USA; and Synopsys, Inc, USA.


\bibliography{report}   
\bibliographystyle{spiejour}   


\vspace{2ex}\noindent\textbf{Mitsuki~Hayashida} is a master's student at Tokyo University of Science in Japan. He received his BS degree in physics from Tokyo University of Science in 2019. His current research interest includes development of X-ray SOI pixel sensors for X-ray astronomical satellites.


\vspace{1ex}
\noindent Biographies and photographs of the other authors are not available.

\listoffigures

\end{spacing}
\end{document}